**Using modified intention-to-treat as a principal stratum estimator for failure to initiate treatment**

Brennan C Kahan[1], Ian R White[1], Mark Edwards[2], Michael O Harhay[3, 4]

1 MRC Clinical Trials Unit at UCL, London, UK

2 Department of Anaesthesia, University Hospital Southampton NHS Foundation Trust, Southampton, UK

3 Clinical Trials Methods and Outcomes Lab, PAIR (Palliative and Advanced Illness Research) Center, Perelman School of Medicine, University of Pennsylvania, Philadelphia, PA, USA

4 Department of Biostatistics, Epidemiology, and Informatics, Perelman School of Medicine, University of Pennsylvania, Philadelphia, PA, USA

**Correspondence to**: Brennan Kahan, b.kahan@ucl.ac.uk

**Funding**

BCK and IRW are funded by the UK MRC, grants MC_UU_00004/07 and MC_UU_00004/09.

**Word count: 3130**

**Abstract**

**Background**

A common intercurrent event affecting many trials is when some participants do not begin their assigned treatment. For example, in a double-blind drug trial, some participants may not receive any dose of study medication. Many trials use a "modified intention-to-treat" (mITT) approach, whereby participants who do not initiate treatment are excluded from the analysis. However, it is not clear (i) the estimand being targeted by such an approach; and (ii) the assumptions necessary for such an approach to be unbiased.

**Methods**

Using potential outcome notation, we demonstrate that a mITT analysis which excludes participants who do not begin treatment is estimating a *principal stratum* estimand (i.e. the treatment effect in the subpopulation of participants who would begin treatment, regardless of which arm they were assigned to). The mITT estimator is unbiased for the principal stratum estimand under the assumption that the intercurrent event is not affected by the assigned treatment arm, that is, participants who initiate treatment in one arm would also do so in the other arm (i.e. if someone began the intervention, they would also have begun the control, and vice versa).

**Results**

We identify two key criteria in determining whether the mITT estimator is likely to be unbiased: first, we must be able to measure the participants in each treatment arm who experience the intercurrent event, and second, the assumption that treatment allocation will not affect whether the participant begins treatment must be reasonable. Most double-blind trials will satisfy these criteria, as the decision to start treatment cannot be influenced by the allocation, and we provide an example of an open-label trial where these criteria are likely to be satisfied as well, implying that a mITT analysis which excludes participants who do not begin treatment is an unbiased estimator for the principal stratum effect in these settings. We also give two examples where these criteria will not be satisfied (one comparing an active intervention vs. usual care, where we cannot identify which usual care participants would have initiated the active intervention, and another comparing two active interventions in an unblinded manner, where knowledge of the assigned treatment arm may affect the participant's choice to begin or not), implying that a mITT estimator will be biased in these settings.

**Conclusions**

A modified intention-to-treat analysis which excludes participants who do not begin treatment can be an unbiased estimator for the principal stratum estimand. Our framework can help identify when the assumptions for unbiasedness are likely to hold, and thus whether modified intention-to-treat is appropriate or not.

**Background**

A common intercurrent event affecting many randomised controlled trials (RCTs) is when some participants do not begin the treatment they were randomised to[1]. For example, the FLO-ELA trial (ISRCTN 14729158) compared the use of a cardiac output monitor against usual care in determining when and how much fluid to give to participants undergoing emergency laparotomy. Due to practical considerations, there is usually a small delay between randomisation and the start of surgery, and a small number of participants had their surgery cancelled after randomisation, either because they deteriorated and were too unwell to undergo surgery, or because the underlying issue resolved itself. Here, a cancelled surgery precludes the participant from initiating treatment, as fluids are no longer relevant for participants who do not undergo their planned surgery. Similar issues may arise in other trials, where participants do not begin their study treatment, or do not receive any dose of study drug.

Many trials use a "modified intention-to-treat" (mITT) approach to handle these intercurrent events in the analysis, whereby participants who do not initiate treatment are excluded (i.e. the analysis is performed only on the participants who initiate treatment)[1-4]. However, this approach has been criticised on the basis that post-randomisation exclusions can induce bias[1]. Further, it is not clear exactly *what* treatment effect is being estimated by mITT in this context. With the recent publication of the ICH-E9(R1) addendum, there is growing recognition for the need to first define the target estimand (the treatment effect we want to estimate), and then choose an estimator which is aligned to that estimand under plausible assumptions[5].

Given the widespread use of modified intention-to-treat in practice, there is urgent need to the target estimand for such an approach, as well as the settings in which it will be unbiased. The purpose of this paper is therefore to (i) demonstrate that an mITT estimator which excludes participants who do not begin treatment targets a *principal stratum* estimand; (ii) provide the assumptions required for the mITT estimator to be unbiased for the principal stratum estimand; and

(iii) provide guidance on how to identify whether the mITT estimator is likely to be unbiased for a given trial.

**Methods**

We begin by defining the principal stratum estimand, and then show how a mITT estimator can correspond to this estimand under certain assumptions. We then provide a simple framework for determining whether the mITT estimator is likely to be unbiased in practice for a given trial.

*Principal stratum estimand*

A principal stratum strategy to handle intercurrent events implies interest lies in the treatment effect in the subpopulation of participants who would (or would not) experience the intercurrent event[5-9]. Here, we define the principal stratum estimand in terms of the intercurrent event of interest, which we refer to as "failure to initiate treatment". Depending on the trial, this could be defined to include any cause of initiation failure (e.g. in a placebo-controlled drug trial where participants do not begin their course of study treatment, regardless of reason), or be defined based on a single cause of initiation failure, with other causes handled using alternate strategies (e.g. in FLO-ELA, where surgery cancellation precludes treatment initiation, but participants may also not initiate for other reasons).

The estimand can be written as:

$$E\bigl(Y^{(Z=1)} - Y^{(Z=0)} | I^{(Z=1)} = I^{(Z=0)} = 1\bigr) \quad (1)$$

where $Z$ denotes treatment (0=control, 1=intervention), $Y^{(Z=1)}$ and $Y^{(Z=0)}$ denote the participant's potential outcome under intervention and control respectively, and $I^{(Z=1)}$ and $I^{(Z=0)}$ are indicator variables denoting whether the participant would initiate treatment under intervention and control respectively (where 0 denotes they would not initiate treatment, and 1 denotes they would). This estimand implies the treatment effect in the subpopulation of participants for whom $I^{(Z=1)} = I^{(Z=0)} = 1$ (i.e. who would initiate treatment under either treatment arm) is of interest.

We note that the definition above primarily focusses on how the intercurrent event "failure to initiate treatment" is handled; the other aspects which make up the estimand (population, treatment conditions, endpoint, summary measure, and handling of other intercurrent events) would also need to be specified in order for the estimand to be complete [5].

*Modified intention-to-treat estimator*

The mITT estimator, which excludes participants who do not begin treatment from the analysis population, can be written as:

$$E(Y|Z = 1, I_{Z=1} = 1) - E(Y|Z = 0, I_{Z=0} = 1) \quad (2)$$

where $Y$ is the participant's observed outcome and $I_{Z=1}$ is an indicator denoting whether the participant initiated treatment under $Z = 1$ and similarly for $I_{Z=0}$. Hence, this estimator compares the set of participants in the intervention group initiated the intervention treatment against the set of participants in the control group who initiated the control treatment.

To evaluate the properties of this estimator we begin by outlining the four principal strata defined by the intercurrent event "failure to initiate treatment" (Table 1)[9].

The "Always initiators" are participants who would initiate treatment in either treatment group (i.e. for whom $I^{(Z=1)} = I^{(Z=0)} = 1$), and the "never initiators" are participants who would not initiate treatment regardless of treatment group (i.e. $I^{(Z=1)} = I^{(Z=0)} = 0$). The "intervention initiators" are participants who would only initiate treatment in the intervention arm (i.e. $I^{(Z=1)} = 1$ and $I^{(Z=0)} = 0$), and the "control initiators" are participants who only initiate treatment if they are in the control group (i.e. $I^{(Z=1)} = 0$ and $I^{(Z=0)} = 1$).

In a RCT, randomisation can be seen to balance the treatment groups within each stratum in expectation. Hence, if we knew the principal stratum membership of each participant, we could perform the analysis in the stratum of interest and obtain an unbiased estimator based on a randomised comparison. However, in practice we do not know the principal stratum membership, as this requires knowledge of participants' initiation status under both treatment conditions, but we can only observe a participant's initiation status under the treatment arm to which they were randomised. For instance, if a participant allocated to the intervention initiated treatment, we have no way of telling whether they are an "Always initiator" or an "Intervention initiator".

We can see from Table 1 that the estimator described in (2) is not based on a randomised comparison, as it compares different principal strata between the treatment arms (i.e. the analysis population differs between treatment groups)[10]. The intervention arm includes the "always initiators" and the "intervention initiators", while the control group consists of the "always initiators" and the "control initiators". Hence, if these two strata (the "intervention initiators" and "control initiators") differ in terms of their potential outcomes, this estimator will be biased.

*Assumptions for unbiasedness*

Ensuring the mITT estimator in (2) is unbiased for the principal stratum estimand in (1) requires a way of identifying the principal stratum membership of participants. This is possible under the assumption that there are no "intervention initiators" and no "control initiators" in the trial, i.e. that *if someone does not initiate treatment in one arm, they would also not initiate treatment in the other arm (and, conversely, if someone does initiate treatment in one arm, they would also initiate treatment in the other arm).*

This assumption implies there are only two principal strata in the trial, "never initiators" and "always initiators"; hence, if we observe a participant's initiation status in one treatment arm, we can infer their initiation status had they been allocated to the alternative treatment, and identify the principal stratum to which they belong (i.e. if we observe $I_{Z=1} = 1$ then we know that $I_{Z=0} = 1$ as well, and hence the participant belongs to the "always initiators" group, and vice versa).

Then, the mITT estimator in (2), which only includes participants who initiated treatment in their assigned arm, will in fact be a comparison within the "always initiators" stratum, which, as discussed above, is based on a randomised comparison (Table 2). As such, this estimator will be unbiased for the principal stratum estimand in (1).

This assumption can be partly assessed, on the basis that if it is true, then the proportion of non-initiators should, on average, be the same across randomised groups. Hence, if there are large discrepancies in the proportion of non-initiators between groups, this may provide evidence that this assumption has been violated. However, even if this proportion is the same across groups, this is not a guarantee the assumption is true, hence this assumption cannot be fully tested, and therefore must rely primarily on contextual information around its underlying plausibility.

We demonstrate the unbiasedness of the mITT estimator under the assumptions given above as follows. First, under consistency assumption[11] (which states that if $Z = z$, then $Y = Y^{(Z=z)}$), we can write $Y|Z = 1$ as $Y^{(Z=1)}$ and $Y|Z = 1$ as $Y^{(Z=0)}$, as well as $I_{Z=1}$ as $I^{(Z=1)}$ and similarly for $I_{Z=0}$. Then, using randomisation, the mITT estimator in (2):

$$E(Y|Z = 1, I_{Z=1} = 1) - E(Y|Z = 0, I_{Z=0} = 1)$$

can be written as:

$$= E(Y^{(Z=1)}|I^{(Z=1)} = 1) - E(Y^{(Z=0)}|I^{(Z=0)} = 1)$$

Then, under the assumption that there are no "intervention initiators" and no "control initiators", $I^{(Z=1)} = 1$ implies that $I^{(Z=0)} = 1$, and vice versa for $I^{(Z=0)}$, implying that $I^{(Z=1)} = I^{(Z=0)} = 1$. Then, the estimator becomes:

$$E(Y^{(Z=1)}|I^{(Z=1)} = 1) - E(Y^{(Z=0)}|I^{(Z=0)} = 1)$$

$$= E\bigl(Y^{(Z=1)}|I^{(Z=1)} = I^{(Z=0)} = 1\bigr) - E\bigl(Y^{(Z=0)}|I^{(Z=1)} = I^{(Z=0)} = 1\bigr)$$

$$= E\bigl(Y^{(Z=1)} - Y^{(Z=0)}|I^{(Z=1)} = I^{(Z=0)} = 1\bigr)$$

which is the principal stratum estimand in (1).

We note that this estimator is unbiased regardless of what happens to "never initiators" (e.g. regardless of whether they receive no treatment at all, or receive a non-trial treatment instead).

*Determining whether the mITT estimator is appropriate*

Two key factors can be used to determine whether the "Always initiators" principal stratum can be identified, thus ensuring the mITT estimator in (2) is appropriate. First, the intercurrent event of interest must be identifiable in each treatment arm (i.e. we must be able to measure which participants in each treatment group experience the intercurrent event).

Secondly, we must be able to assume that the occurrence of the intercurrent event is not affected by treatment allocation (i.e. that if the intercurrent event occurs for a participant under one treatment condition, it would also occur under the other treatment condition). Together, these two factors justify the assumption of no "intervention initiators" and no "control initiators".

**Results**

Below we give several examples of trials both where the mITT estimator is and is not appropriate (Table 3).

*FLO-ELA*

As described earlier, FLO-ELA is an open-label trial comparing two methods of fluid delivery (cardiac output monitor vs. clinician judgement) in participants undergoing emergency laparotomy. The primary outcome measure is the number of days that participants are alive and out of hospital within 90 days of randomisation. The intercurrent event of interest is the cancellation of surgery, which precludes treatment from being initiated (as fluid delivery is only relevant once participants begin surgery). This is an example of a "cause-specific" initiation failure, as some participants who do undergo surgery may also not initiate treatment for other reasons. We note that ideally randomisation would be done at the point surgery was due to begin, to minimise the number of cancellations, however in practice this is not always feasible due to the time required to prepare the intervention combined with the emergency setting.

In this trial, the mITT estimator is appropriate. First, the intercurrent event can be easily identified and measured in both treatment groups, and secondly, the decision on whether to cancel surgery or not will almost certainly not be affected by the allocated treatment arm. This is because in most cases the relevant decision makers (surgeons) will be unaware of trial group allocation until surgery starts (i.e. at the point the decision is made). Further, the decision not to proceed with surgery has large health implications for the patient and is only undertaken in response to a major change in the patient's clinical condition since surgery was initially planned, and it is implausible that such a fundamental change in patient care would be undertaken on the basis of the planned method of fluid delivery.

*MIST2*

MIST2 was a 2x2 factorial trial evaluating the use of tPA and DNase in participants with pleural infection[12]. We focus on the tPA comparison here. tPA was compared in a double-blind fashion against a matching placebo, so that neither participants nor clinicians were aware of whether participants were receiving tPA or placebo. Treatment was to be given for three days. The intercurrent event of interest is failure to begin treatment (i.e. if participants did not receive any dose of study drug), regardless of the reason.

In this trial, the mITT estimator is appropriate. First, the intercurrent event ("failure to begin study treatment") can be identified and measured in each treatment arm. Secondly, treatment allocation should not influence the occurrence of the intercurrent event, as the trial is double-blinded, so those deciding on whether to initiate treatment will not be aware of whether it is tPA or placebo being initiated.

*COPERS*

The COPERS trial evaluated the use of a group pain management support intervention for participants with chronic musculoskeletal pain[13]. The intervention consisted of a group intervention delivered over three days with a follow-up session at two weeks. The control consisted of usual care. The intercurrent event of interest is failure to begin the group intervention session (i.e. failure to attend any sessions).

Here, the mITT estimator is *not* appropriate, as there is no way of identifying which participants in the control arm would attend any sessions if allocated to the group intervention, and so the principal stratum of interest cannot be identified.

*SWAP*

The SWAP trial evaluated the use of a multi-modal group intervention to reduce weight in obese adults[14]. The intervention consisted of eight group sessions delivered by trained advisors, while the control consisted of four standard weight management sessions delivered by a practice nurse. Due to the nature of the interventions, the trial was open-label, with participants and healthcare professionals aware of the allocated group. The intercurrent event of interest is failure to attend any treatment sessions.

In this trial, the mITT estimator is *not* appropriate. Although the intercurrent event can be identified in each treatment arm, the occurrence of the intercurrent event may be affected by the allocated arm. For instance, due to participants' perceptions about the two treatment arms, they may be more likely to attend the group intervention compared to the standard nurse-led weight management sessions, or vice versa. Hence, the assumption of no "intervention initiators" and no "control initiators" is likely to be false, and the estimator will be based on a non-random comparison, and hence subject to bias.

*Reporting of this estimator*

Given that the mITT estimator in (2) relies on certain assumptions for unbiasedness, its use must be transparently reported so that readers may evaluate whether the assumptions underpinning the study results are reasonable.

To provide full clarity, we suggest the methods be reported according to Box 1. This provides the target of estimation (the estimand), which alerts readers a principal stratum effect is of interest; explicitly states that participants who experience the intercurrent event are excluded from the analysis population, along with the reason for doing so; provides the assumptions required for this estimator to be unbiased; and finally, justifies those assumptions for the specific trial. Together, this provides readers with an understanding of *what* is being estimated, how the estimation procedure works, the assumptions required, as well as the rationale underpinning those assumptions.

**Discussion**

Failure to initiate treatment is a common intercurrent event in randomised trials. A modified intention-to-treat approach is frequently used, whereby participants who do not begin treatment are excluded from the analysis, however it is not clear what estimand is being targeted by such an approach, or the assumptions required for unbiasedness. In this paper, we show that the mITT estimator which excludes participants who do not begin treatment targets a principal stratum estimand, and is unbiased under the assumption that if a participant experiences the intercurrent event under one treatment condition, they would also experience it under the other conditions.

This assumption is likely to be fulfilled in many settings, including most double-blind trials, where the decision to initiate treatment will not be affected by the treatment arm. This assumption may also be plausible in many open-label trials as well; for instance, in FLO-ELA, there is strong justification for expecting cancellation of surgery not to be affected by treatment group. However, these assumptions are not always appropriate, and when they are not, the mITT estimator will be biased. The results in this paper can help investigators determine whether a mITT approach is appropriate for their trial. When the underlying assumptions cannot be justified, alternative methods to estimate the principal stratum estimand can be used (see references 6 and 7 for an overview of approaches to estimate principal stratum effects in more challenging settings).

In addition to determining whether the mITT estimator will be unbiased, investigators also need to decide whether a principal stratum estimand is appropriate to use. This decision will depend on the specific aims of each individual trial, however we argue that for failure to initiate treatment, a principal stratum strategy often is the most clinically relevant choice. For example, in the FLO-ELA trial, cancellation of surgery renders the participant no longer part of the population of interest (which is participants undergoing emergency laparotomy); hence, a principal stratum approach may be more reflective of the treatment effect in routine clinical practice compared to other strategies, such as treatment policy (which compares two different strategies of fluid delivery during surgery, *regardless of whether participants actually undergo surgery*, which is clearly not of direct clinical interest).

Despite the appeal of the principal stratum estimand, it has rarely been used in practice[15]. This may in part be because in more complex settings principal stratum estimators require both more complex estimation procedures as well as more complex underlying assumptions which cannot always be verified based on contextual information about the trial. One benefit of our approach is that it is far more straightforward, both conceptually and to estimate, and so may be simpler to implement in practice.

Modified ITT analyses have been criticised on the basis of ambiguity, i.e. the term has been used by different researchers to mean different things[2-4, 16]. If using this term to describe the estimator, we recommend it be described as in Box 1 (i.e. that it is explicitly stated which intercurrent events are excluded), to avoid potential misunderstandings. It is also not always clear which estimand such mITT analyses have targeted in the past. The results in this paper can serve to clarify the target of estimation for these analyses, as well as the assumptions required to justify their validity.

Although this paper has focussed on the intercurrent event "failure to initiate treatment", the results could also apply to other intercurrent events where the assumptions discussed here are likely to be fulfilled. For instance, in trials where mortality is an intercurrent event but the treatments are known to not affect mortality (for instance, in certain palliative care trials), then (if clinically appropriate) the estimator described in (2) could also be used.

**Conclusion**

A modified intention-to-treat analysis which excludes participants who do not begin treatment can be an unbiased estimator for the principal stratum estimand. Our framework can help identify when the assumptions for unbiasedness are likely to hold, and thus whether modified intention-to-treat is appropriate or not.

**Table 1: The four principal strata defined by the intercurrent event "failure to initiate treatment".**

The mITT estimator excluding participants who do not initiate treatment compares the shaded cells in the intervention column against the shaded cells in the control column.

|  | Initiate treatment? | |
| --- | --- | --- |
|  | Control ($Z = 0$) | Intervention ($Z = 1$) |
| Always initiators | Yes | Yes |
| Intervention initiators | No | Yes |
| Control initiators | Yes | No |
| Never initiators | No | No |

**Table 2: The mITT principal stratum estimator under the assumption of no "intervention initiators" and no "control initiators".** The estimator compares the shaded cells in the intervention column against the shaded cells in the control column. Grey indicates that these strata do not exist.

|  | Initiate treatment? | |
| --- | --- | --- |
|  | Control ($Z = 0$) | Intervention ($Z = 1$) |
| Always initiators | Yes | Yes |
| ~~Intervention initiators~~ | ~~No~~ | ~~Yes~~ |
| ~~Control initiators~~ | ~~Yes~~ | ~~No~~ |
| Never initiators | No | No |

**Table 3 – Examples of trials where the mITT principal stratum estimator both is and is not appropriate**

| Trial | Description | Intercurrent event | mITT principal stratum estimator appropriate? | Rationale |
|---|---|---|---|---|
| FLO-ELA | An open-label trial comparing use of a cardiac output monitor vs. clinical judgement for fluid delivery in participants undergoing emergency laparotomy | Cancellation of surgery | Yes | Intercurrent event ("surgery cancellation") can be identified in both treatment arms, and the decision to cancel surgery is unlikely to be affected by the arm that participants are allocated to |
| MIST2 | A blinded trial comparing tPA vs. placebo in participants with pleural infection | Failure to receive any dose of study treatment | Yes | Intercurrent event ("failure to initiate treatment") can be identified in both arms, and will not be affected by the arm that participants are allocated to due to blinding |
| COPERS | An open-label trial comparing a group pain management support intervention vs. usual care in participants with chronic musculoskeletal pain | Failure to attend any sessions of the group intervention | No | There is no way of identifying which participants in the usual care arm would attend the group intervention, and hence no way of identifying the principal stratum of interest |
| SWAP | An open-label trial comparing a group intervention vs. a standard weight management session delivered by practice nurses in obese adults | Failure to attend any sessions of allocated treatment | No | Attendance may be affected by the allocated treatment arm (e.g. participants may be more likely to attend the group intervention compared to the standard weight management sessions, or vice versa), and thus the estimator will be based on a non-randomised comparison |

**Box 1 – Reporting recommendations for using the mITT principal stratum estimator.**

Recommendations are based on the mITT principal stratum estimator defined in formula (2) for the intercurrent event "failure to initiate treatment".

- In the estimand definition, state that a principal stratum approach is being used for the intercurrent event
- In the description of the estimator, state that participants who experience the intercurrent event are excluded from the analysis population as a way to estimate the principal stratum effect
- State the assumptions required for this estimator to be unbiased (i.e. no "intervention initiators" and no "control initiators")
- Provide an explanation for why the assumptions above are justified in the trial

**Funding**

BCK, and IRW are funded by the UK MRC, grants MC_UU_00004/07 and MC_UU_00004/09.

**Contributions**

BCK wrote the first draft of the manuscript. IRW, ME, and MH revised the manuscript. All authors read and approved the final manuscript.

**Declaration of Conflicting Interests**

The Authors declare that there is no conflict of interest.

**Data availability**

This article contains no data.